\documentclass[plb,showpacs,superscriptaddress,preprintnumbers,amsmath]{revtex4}
\usepackage[dvips,final]{graphicx}
  \usepackage{amssymb}
   \usepackage{amsmath}
    \usepackage{epsfig}
     \usepackage{bm}
      \usepackage{pifont}

\newcommand{\vecc}[1]{\mbox{\boldmath $#1$}}

\def\({\left(}
\def\[{\left[}
\def\){\right)}
\def\]{\right]}

\begin{document}
\preprint{\hbox{RUB-TPII-09/2010}}

\title{On correlations in high-energy hadronic processes and
       the CMS ridge: \\
       A manifestation of quantum entanglement?}
\author{I.~O.~Cherednikov}
\email{igor.cherednikov@ua.ac.be}
\affiliation{Departement Fysica, Universiteit Antwerpen, B-2020 Antwerpen, Belgium\\
 and \\
             Bogoliubov Laboratory of Theoretical Physics,
             JINR,
             141980 Dubna, Russia}
\author{N.~G.~Stefanis}
\email{stefanis@tp2.ruhr-uni-bochum.de}
\affiliation{Institut f\"{u}r Theoretische Physik II,
             Ruhr-Universit\"{a}t Bochum,
             D-44780 Bochum, Germany\\}

\date{\today}

\begin{abstract}
We discuss the possibility of quantum entanglement
for pairs of charged particles produced in high-energy
$pp$-collisions at the LHC.
Using a framework of interacting Wilson lines, we calculate 2-D
and 1-D two-particle angular correlation functions in terms of
the differences of the pseudorapidities and azimuthal angles of
the produced particles.
The calculated near-side angular correlation shows a localized
maximum around $\Delta\phi \approx 0$, though it is less pronounced
compared to the peak observed by the CMS Collaboration.
We argue that this soft correlation is universal and insensitive to
the specific properties of the matter (quark-gluon plasma,
QCD vacuum, etc.) used to describe hadronic states---though such
properties can be included to further improve the results.
\end{abstract}


\maketitle

\section{Introduction}
\label{sec:intro}

The CMS collaboration at the Large Hadron Collider (LHC) of CERN
observed last year \cite{CMS_2010_ridge} significant two-particle
angular correlations of charged particles, produced in proton-proton
($pp$) collisions at a center-of-mass energy $\sqrt s = 7 \ {\rm GeV}$.
These correlations are of long-range nature in the pseudorapidity
difference $2.0 < |\Delta \eta| < 4.8$
and appear on a three-dimensional (3-D) plot of the
$\Delta\eta-\Delta\phi$ correlation function as a clearly seen
ridge-like structure on the near-side $\Delta \phi \approx 0$.
Here $\Delta\eta=\eta_{1}-\eta_{2}$ is the difference in the
pseudorapidity $\eta = -\ln [\tan (\theta/2)]$ between the two
produced particles, where $\theta$ is the polar angle relative to
the beam axis, and $\Delta \phi=\phi_{1}-\phi_{2}$ is the difference
in their azimuthal angle.
The ridge becomes most pronounced for intermediate transverse momenta
$1 \ {\rm GeV} < p_\perp < 3 \ {\rm GeV}$
and in events with an average charged-particle multiplicity of
at least $N\approx 90$, whereas simulations using Monte Carlo (MC)
models do not predict such an effect---independent of multiplicity and
transverse momentum.
No evidence for such near-side long-range correlations has been found
in the $pp$ or $p\bar{p}$ data before, though the novel ridge structure
is reminiscent of correlations observed in nuclear collisions at the
Relativistic Heavy Ion Collider (RHIC) at Brookhaven National Laboratory
\cite{PHOBOS07,PHOBOS10,STAR10}.
However, these correlations were assumed to arise from interactions of
the produced particles with the medium and from collective effects in
the \textit{initial-state} interactions of the nuclei.

Adopting the notations of \cite{Corr_Def,PHOBOS07}, the two-particle
correlation function can be defined as follows:
\begin{equation}
  R (\eta_1, \phi_1, \eta_2, \phi_2)
=
  \Bigg\langle \!\! (\langle N\rangle -1) \!
               \( \frac{\rho_{II} (\eta_1, \phi_1, \eta_2, \phi_2)}
               {\rho_I (\eta_1, \phi_1)\rho_I (\eta_2, \phi_2)} - 1
               \) \!
  \!\! \Bigg\rangle \ ,
\label{eq:corr_fun}
\end{equation}
where
$\rho_I (\eta, \phi)$
and
$\rho_{II} (\eta_1, \phi_1, \eta_2, \phi_2)$
denote single and double particle distributions in the final state,
respectively \cite{Corr_Def}.
In what follows, we will express the correlation function in
terms of {\it parton} distributions which differ, in general, from
those of real particles (hadrons).
This approximation is justified because the correlations under
consideration emerge long before the hadronization starts, so that the
process can be adequately described in terms of partons.
In the context of the CMS experiment,
$\rho_{II}(\eta_1, \phi_1, \eta_2, \phi_2)$ resembles
$S_N$, i.e., the signal distribution, whereas the product of the
single parton distributions plays the role of the random background
distribution $B_N$---cf.\ Eqs.\ (4.2) and (4.3) in
\cite{CMS_2010_ridge}.
Moreover, $\langle N\rangle$ is the number of tracks per event averaged
over the multiplicity bin and $R(\Delta\eta, \Delta\phi)$ is determined
by averaging over multiplicity bins and over
$(\eta_1 + \eta_2)$ and $(\phi_1+\phi_2)$---see \cite{Corr_Def}.
These quantities will be specified and discussed in detail within our
theoretical framework later.

A wide pseudorapidity range of correlated charged-particle pairs
indicates that the source of the correlation has to be primordial,
in the sense that it emerges coincidentally with the particle
production, but before hadronization occurs
(see, e.g., \cite{CGC_2009, CGC_2010}).
Theoretical attempts to dig out the physical origin of the ridge
structure in the high-multiplicity $pp$ data observed by CMS
mainly concentrate on the properties of a particular state of
matter---the Color Glass Condensate (CGC)---that was created during
the high-energy collision
(see, for instance, Refs.\ \cite{Shuryak, CGC_2010, IHEP10}).
Other authors claim that the observed ridge may be the manifestation
of an azimuthal quadrupole \cite{TK10}, or the result of reflective
scattering effects in multiparticle production processes
\cite{IHEP11}.
Further references concerning the ridge formation mechanisms
can be found in \cite{CMS_TH_new}, where the CMS Collaboration have
updated their analysis to include the full 2010 statistics.

We are going to approach the emergence of the ridge phenomenon from
another side which has echoes of our previous works on the gauge
invariance of transverse-momentum-dependent (TMD) parton
distribution functions (PDF)s (see \cite{CS_all, CKS_2010}, and
\cite{SC09} for reviews).
To be precise, we intend to discuss effects ensuing from nonlocal
path-dependent interactions---the latter naturally arising in the
gauge-invariant description of high-energy processes with charged
particles in terms of Mandelstam fields \cite{Man62}.
We will argue in what follows that the inclusion of semi-infinite
gauge links (Wilson lines) gives rise to a phase that in some sense
is akin to the intrinsic Coulomb phase found by Jakob and Stefanis (JS)
\cite{JS91} in QED by employing Mandelstam's gauge-invariant
path-dependent formalism.
The origin of the intrinsic Coulomb phase can be considered as a
residual effect of the primordial separation of the particle from its
oppositely charged counterpart and has no bearing on external charge
distributions.

Adopting this viewpoint, we will work out a scheme which will enable
us to calculate the leading-order contribution to the two-dimensional
(2-D) $\Delta\eta-\Delta\phi$ correlation function of interacting
Wilson lines.
The justification for such a treatment is provided by the fact that
Wilson lines can be recast in the form of particle worldlines in the
context of first-quantized theories \cite{Pol79}.
Therefore, this scheme lends itself to the actual experimental
situation in which charged particles are emitted in the $pp$ collisions
at a very high center of mass energy $\sqrt{s}$ of 0.9, 2.36, and 7
TeV, forming highly collimated jets before they finally hadronize.
The key idea here is that as the number of such particle tracks
increases, several pairs of them are created in unison (with regard to
their separation in $\Delta\eta$ and $\Delta\phi$), so that the
quantum mechanism of entanglement applies.
Note in this context that it is not necessary to consider these pairs
as being formed by particles with opposite polarities, as discussed by
JS in the special case of the QED intrinsic Coulomb phase.
In fact, the CMS data analysis \cite{CMS_2010_ridge} has shown that the
results for like-sign and unlike-sign pairs agree with each other
within uncertainties.
Also our calculation does not depend on the sign of the correlated
pairs.
As we will see below, the obtained result shows a clear and
significant ridge-like structure in the plane of the differences in
pseudorapidity and azimuthal angle between the interacting Wilson
lines, which occurs around
$\Delta\phi\approx 0$ and extends over a wide range of $\Delta\eta$,
albeit the near-side peak is less prominent compared to that
observed by CMS.
The rest of the paper is organized as follows.
In the next section, we recall the main features of the intrinsic
Coulomb phase in QED in order to sketch the main idea underlying
the present application.
Our theoretical approach to understand the ridge structure and the
comparison with the experimental findings of the CMS Collaboration
are presented in Sec.\ \ref{sec:ridge}.
In Sec. \ref{sec:summary}, we give a summary of our results and
discuss them further.
Finally, in Sec.\ \ref{sec:concl} we draw our conclusions.

\section{Intrinsic Coulomb phase in QED}
\label{sec:coulomb}

It is worth indulging in a quick detour into the nature of the
``intrinsic'' Coulomb phase \cite{JS91} which is an extreme case
of quantum entanglement because it persists even for infinitely
separated particles.

Within the JS path-dependent approach, charged fields are always
accompanied by a soft-photon cloud in terms of a non-integrable
phase factor which contains a timelike straight line stretching to
infinity.
In the presence of other charges, the charged field acquires a
relativistic Coulomb phase, in accordance with the results obtained
earlier by Kulish and Faddeev \cite{KF70}, and independently by
Zwanziger \cite{Zwa_all}.
But, it also yields an additional phase that does not depend on
external charge distributions and remains different from zero, even
if the conventional Coulomb phase vanishes---hence, the name
``intrinsic''.
The origin of this phase was ascribed by JS to the asymptotic
interaction of the charged particle with its oppositely charged
counterpart that was removed ``behind the moon'' instantaneously
after their common primordial creation.
The mechanism of the particle creation, as well as that of their
separation, is irrelevant for these considerations.
What counts is that they have been created in common, i.e., entangled.
Speaking in more technical terms, the intrinsic Coulomb
phase is acquired during the parallel transport of the
charged field along a timelike straight line from infinity to the
point of interaction with the photon field and is absent in the
local approach, i.e., for charged fields joined by a (gauge)
connector (see, e.g., \cite{Ste83} for more details).
It is different from zero only for a Mandelstam field, which has
its own gauge contour attached to it, and keeps track of its full
history since its primordial entangled creation with its antiparticle.
Let us also mention that the existence of a balancing charge ``behind
the moon'' was postulated before by several authors---see \cite{JS91}
for related references---in an attempt to restore the Lorentz
covariance of the charged sector of QED.
Therefore, the discussed mechanism of phase entanglement originates
from quite general features of gauge field theory and is insensitive
to the specific properties of the particular process under
consideration.
We want to make it clear at this point that we are not advocating a
violation of the principle of locality: nothing can cross a void
linking cause to effect---except a phase.

The simplest non-Abelian generalization of the intrinsic Coulomb
phase for the quark fields reads (in leading order of the strong
coupling constant $\alpha_s$ indicated by the superscript $(1)$)
\begin{equation}
  \Phi^{(1)}(\Gamma)
=
  \ - 4\pi
  \int_{{\Gamma}}\! dx_\mu  \int d^4 z
  D_{ab}^{\mu\nu} (x-z)t^a  j_\nu^b (z) \ ,
\label{eq:phase}
\end{equation}
where $D_{ab}^{\mu\nu}$ is the (free) gluon propagator, and the current
$ j_\nu^b (z)
=
  t^b v_\nu \int_{\Gamma}
  d\tau \delta^{(4)}(z - v\tau) $
describes the residual effects of the oppositely charged counterpart in
a semiclassical approximation.
The integration path $\Gamma$ is the contour defined by the kinematics
of the process and can have a complicated structure venturing
off-the-light-cone in the transverse direction \cite{CS_all,CS10}.
Note that the phase ({\ref{eq:phase}}) can be considered as the
leading non-trivial term of the expansion of a non-Abelian path-ordered
exponential \cite{KR87} in the ground state (abbreviated by GS), i.e.,
\begin{widetext}
\begin{equation}
  \Phi (\Gamma)
=
  \exp \left[\sum_{n=1}^{\infty} \alpha_s^n \Phi^{(n)} (\Gamma)
       \right]
=
  \left\langle {\rm GS}
  \left| {\cal P}
  \exp\Big[ig \int_{\Gamma}\! d\zeta^\mu
           \ t^a A^a_\mu (\zeta)
      \Big]
  \right|{\rm GS}
  \right\rangle  \ .
\label{eq:phase-2}
\end{equation}
\end{widetext}
As we shall show shortly, the path dependence of the phase factor in
({\ref{eq:phase-2}}) becomes crucial in the application of the
Mandelstam formalism in studying semi-inclusive processes in QCD.
\bigskip
\bigskip

\section{CMS ridge---theoretical considerations}
\label{sec:ridge}

The appearance of long-range correlations in nuclear collisions
at RHIC \cite{RHIC05-10} and (very recently) in $pp$-collisions
at the LHC \cite{CMS_2010_ridge} has been studied within the
color-glass-condensate approach in \cite{CGC_2009, CGC_2010}
(see also the more recent papers \cite{Shuryak, IHEP10}).
In our study we refrain from making any assumptions about the
structure of the quark-gluon (QG) matter just-before/during/just-after
the proton-proton collisions, and appeal instead to a field theoretic
description of the particle tracks in terms of path-dependent
exponentials along the lines exposed above.
Nevertheless, the properties of the QG-matter can be taken into account
within our approach by means of a specification of the ground state
entering the evaluation of matrix elements with Wilson-line
exponentials.

To make qualitative estimates, we will now expose our working method
to treat the semi-inclusive process of the two-particle production
in the proton-proton collision in more mathematical terms and start
by writing
\begin{equation}
  h(p_1) + h(p_2) \ \to \
  y(p_3) + y(p_4) + {\cal X} \ .
\label{eq:p-p-collision}
\end{equation}
In what follows, we try to understand this process from the point of
view of the QCD factorized cross-section, our goal being to figure
out what part(s) of this factorized expression is (are) responsible
for the above-mentioned ridge-like structure.
Keep in mind that the observed correlations have to be nonlocal and
primordial so that our task consists in identifying which part in the
factorized expression can serve as the root cause of such a phenomenon.
The (symbolic) factorization formula reads
\begin{equation}
  {d\sigma}
\sim
  H \otimes \tilde{\cal F}_{\rm D_1}
    \otimes \tilde{\cal F}_{\rm D_2}
    \otimes \tilde{\cal F}_{\rm F_1}
    \otimes \tilde{\cal F}_{\rm F_1} + {\rm power \ suppressed \ terms} \ ,
\label{eq:factorization_trial}
\end{equation}
while a detailed discussion of this expression and references can be
found, for instance, in
Refs.\ \cite{S79, Col03, JMY04, BR05, CRS07}.
Problems related to the factorization and universality of TMD
PDFs, etc., are considered in Refs.\ \cite{BM07, CQ07, RM2010}.
Note that these issues are not crucial for our discussion.
We will demonstrate below that the correlations under consideration
can be described by the interaction of Wilson lines without explicit
reference to the factorization formula.
The acronyms used in the above equation and below have the
following meaning:
PFF---parton fragmentation function;
DF---distribution function;
FF---fragmentation function,
so that the subscripts ${\rm D, F}$ are labels for `distribution'
and `fragmentation', respectively.
Note that the right place to take into account TMD PDFs is the
moderate $p_\perp$ region
$1 \ {\rm GeV/c} < p_\perp < 3 \ {\rm GeV/c}$,
where the intrinsic transverse momenta of the partons become relevant,
while at larger $p_\perp$ the transverse momentum is mostly
generated by perturbative hard-gluon exchanges.

Given these circumstances, one has to dig out the origin of the
long-range primordial correlations employing the above factorization
theorem.
We will argue that the observed two-particle correlation has its
origin in the soft factors which are indispensable for the
renormalization of the TMD PDFs \cite{CS_all}.
Each DF (or FF) has its own soft part, so that the above factorization
becomes
\begin{eqnarray}
  {d\sigma}
\sim
  H \otimes [{\cal F}_{\rm D_1}
  \cdot S_{\rm D_1}]
\otimes
       [{\cal F}_{\rm D_2}
       \cdot S_{\rm D_2} ]
\otimes
       [{\cal F}_{\rm F_1}
       \cdot S_{\rm F_1} ]
\otimes
       [{\cal F}_{\rm F_1}
       \cdot S_{\rm F_2}]
       + {\rm power \ suppressed \ terms} \ .
\label{eq:factorization_soft}
\end{eqnarray}
The soft terms can be properly defined in terms of expectation
values of Wilson-line exponentials evaluated along particular
(integration) gauge contours, specified in
Refs.\ \cite{CS_all, CKS_2010}, which enter expressions like
\begin{equation}
  S_{\rm \alpha}
=
  \left\langle {\rm GS}
  \left| {\cal P}
  \exp\Big[ig \int_{\Gamma}\! d\zeta^\mu
           \ t^a A^a_\mu (\zeta)
      \Big]
  \right|{\rm GS}
  \right\rangle \ ,
\label{eq:soft_part}
\end{equation}
with the subscript $\alpha$ labeling the various factors in
Eq.\ (\ref{eq:factorization_soft}).
These soft factors serve to eliminate overlapping divergences
originating from interrelated short and large-scale effects that cannot
be cured by (dimensional) regularization when calculating correlators
of contour-dependent quark and gluon operators.
It can be linked to a cusp in the gauge contour at light-cone
infinity and gives rise to a non-integrable phase factor \cite{CS_all}.

\begin{figure}[h]
\centering
\includegraphics[scale=0.60,angle=90]{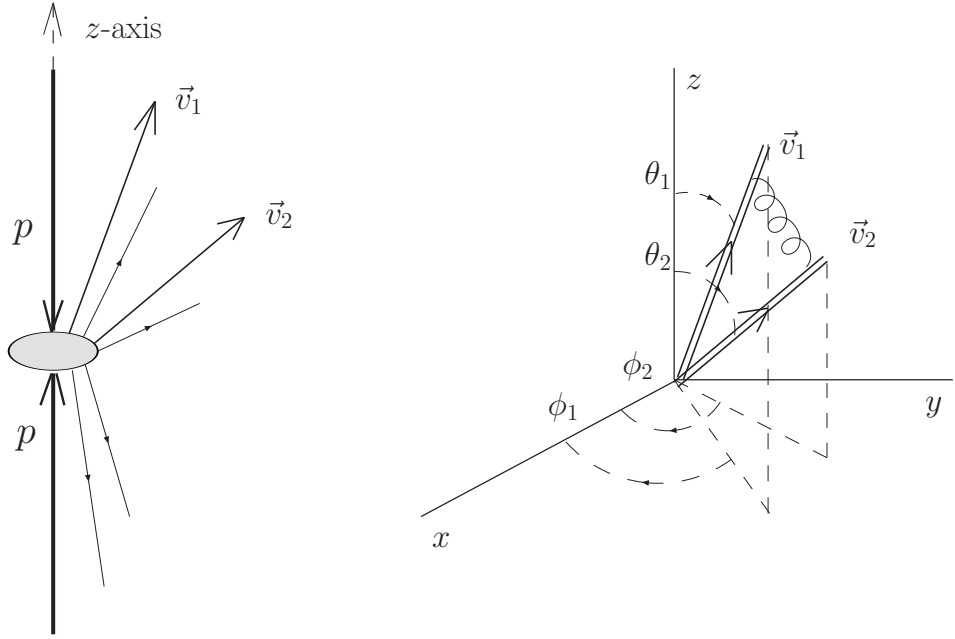}~~
\caption{Left panel: Visualization of the $pp$-collision along
the $z$ axis, with a multiparticle final state and two
produced particles (propagating along the vectors
$\vec v_1$ and $\vec v_2$) being detected and analyzed.
Right panel:
Schematic illustration of interacting soft factors
(Wilson lines)---denoted by double lines---which enter the
fragmentation functions of two particles with the momenta
$p_1$ and $p_2$.
The polar angles $\theta_{1,2}$ and the azimuthal angles $\phi_{1,2}$
correspond to the experimental situation in the CMS experiment,
with the $z-$axis pointing along the beam direction.
Only the one-gluon exchanges are shown which give rise to non-trivial
angle-dependent contributions.
\label{fig:wilson}}
\end{figure}

To implement our considerations, we define the following correlation
function in terms of the soft factors of the fragmentation
functions (the latter describing the hadronization process),
considering it as a simplified version of the true correlation
function given by (\ref{eq:corr_fun}):
\begin{equation}
  K_{\rm soft} (\eta_1, \phi_1; \eta_2, \phi_2)
=
  \frac{S_{\alpha\beta} (\eta_1, \phi_1, \eta_2, \phi_2)}
  {S_{\alpha} (\eta_1, \phi_1) S_{\beta} (\eta_2, \phi_2)} - 1 \ .
\label{eq:corr_soft}
\end{equation}
Taking into account that the soft factors enter the definition
of TMD fragmentation functions multiplicatively, we assume that the
soft correlation function (\ref{eq:corr_soft}) can be used as a
(separately calculable) correction to the full correlation
function $R(\Delta \eta, \Delta \phi)$
that also includes the non-soft contributions from the TMD
PFFs.
Our task is, therefore, to demonstrate that the soft function
(\ref{eq:corr_soft}) gives rise to the long-range rapidity
contributions that cannot be obtained without soft factors.
The soft factors, which describe distribution functions of particle
worldlines (tracks), are defined by
\begin{widetext}
\begin{equation}
   S_{\alpha\beta} (\eta_1, \phi_1, \eta_2, \phi_2)
=
  \left\langle {\rm GS}
  \left| {\cal P}
  \exp\left[ig \int_{\Gamma_{\alpha{\phantom{\beta}}}}\! d\zeta^\mu
           \ t^a A^a_\mu (\zeta)
      \right] \cdot
      {\cal P}
  \exp\left[ig \int_{\Gamma_\beta}\! d\zeta^\mu
           \ t^a A^a_\mu (\zeta)
      \right]
  \right|{\rm GS}
  \right \rangle
\label{eq:product_1}
\end{equation}
and
\begin{equation}
   S_{\alpha}(\eta_1, \phi_1)
   S_{\beta}(\eta_2, \phi_2)
=
  \left\langle {\rm GS}
  \left| {\cal P}
  \exp\left[ig \int_{\Gamma_{\alpha{\phantom{\beta}}}}\! d\zeta^\mu
            t^a A^a_\mu (\zeta)
      \right]
  \right|{\rm GS}
  \right \rangle
  \cdot
  \left\langle {\rm GS}
  \left|
      {\cal P}
  \exp\left[ig \int_{\Gamma_\beta}\! d\zeta^\mu
            t^a A^a_\mu (\zeta)
      \right]
  \right|{\rm GS}
  \right \rangle
\label{eq:product_2}
\end{equation}
\end{widetext}
and are path-dependent quantities because they depend
on the gauge contours.
Referring again to Eq.\ (\ref{eq:factorization_soft}), we note that the
integration contours $\Gamma_\alpha$ and $\Gamma_\beta$ are similar in
structure, but they are separated from each other by the transverse
distance $\vecc b_\perp$ which is equal to the impact parameter of the
collision, as illustrated in the left panel of
Fig.\ \ref{fig:wilson}.
Hence, the two-particle correlation function of experimental tracks
transforms in our scheme into the correlation function of the
worldlines of auxiliary particles moving along the gauge contours
entering the soft factors and can be parameterized in terms of their
separation in pseudorapidity and azimuthal angle.
Pictorially, this theoretical setup of Wilson (or world-) lines looks
like a sheaf of wheat within a small conus of less than about
$10$ degrees corresponding to a $\Delta\eta$ of 2.44.
Hence, pairs of these Wilson lines will be likely produced in
association and may form an entangled quantum state, giving rise to
long-range $\Delta\eta,\Delta\phi$ correlations.
The evaluation of the above expressions below proceeds by calculating
all Feynman graphs which emerge from the perturbative expansion of the
path-ordered exponentials and retaining only the leading-order terms
of order $g^2$.

To start, note that the correlation function (\ref{eq:corr_fun})
depends on the Minkowski angle between the longitudinal parts of the
Wilson lines,
$(p_1^\parallel \cdot p_2^\parallel ) \sim \cosh \Delta Y$,
(where $\Delta Y$ is the difference of the rapidities)
and on the azimuthal angle $\Delta \phi$ between their transverse
parts, i.e.,
$(p_1^\perp \cdot p_2^\perp ) \sim \cos \Delta \phi$
(see the right panel of Fig.\ \ref{fig:wilson} for an illustration).
Thus, azimuthal correlations of the interacting Wilson lines (alias
particle worldlines), entering the soft factors, can be linked
to their transverse-momentum dependence, thus facilitating the
computation.
Then, in leading order $O(g^2)$, the only non-trivial angular
dependence stems from the correlator of two semi-infinite Wilson
lines, evaluated along the rays defined by the
dimensionless four-vectors $v_1$ and $v_2$.
These vectors imitate the momentum vectors of the produced particles
$p_{1,2} = m v_{1,2}$ with $v_{1,2}^2 = 1$.
For the sake of convenience, we fix in Fig.\ \ref{fig:wilson} the
longitudinal and the transverse axes of the mock CMS experimental
setup along the directions $z$ (beam axis) and $x$, respectively.

The momentum vectors of the particles in the final state
are defined as
\begin{eqnarray}
  p_{1,2}^0 = m_T  \cosh  Y_{1,2}, \quad
  p_{1,2}^z = m_T  \sinh  Y_{1,2}, \quad
  p_{1,2}^x = p_\perp  \cos  \phi_{1,2}, \quad
  p_{1,2}^y = p_\perp \sin \phi_{1,2}
\ ,
\label{eq:parameter_rapidity}
\end{eqnarray}
where $Y_{1,2}$ are the rapidities, $\phi_{1,2}$ denote as
before the azimuthal angles of the produced particles,
and $m^2 = m_T^2 - p_\perp^2$.
The above parametrization is more suitable for four-dimensional
worldlines than that based on three-dimensional spherical coordinates
with the pseudorapidity $\eta = - \ln [\tan (\theta/2)]$ and
\begin{equation}
  p_{1,2}^0 = E, \quad
  p_{1,2}^z = p_\perp  \sinh  \eta_{1,2}, \quad
  p_{1,2}^x = p_\perp  \cos  \phi_{1,2}, \quad
  p_{1,2}^y = p_\perp \sin \phi_{1,2}
\ .
\label{eq:parameter_pseudorapidity}
\end{equation}
In the limit $p_\perp \gg m $, for which $\eta \to Y$, both
parameterizations coincide.
This approximation is natural, for instance, for pions, so that
in what follows we will make use of the
``worldline-friendly''
parametrization (\ref{eq:parameter_rapidity}).

Then, by definition, the leading-order non-trivial
angle-dependent term of Eq.\ (\ref{eq:corr_soft}) reads
\begin{eqnarray}
   K^{(1)} (Y_1, Y_2, \phi_1, \phi_2;p_\perp, m_T)
=
  \frac{[1 + (ig)^2  v_1^\mu v_2^\nu
  \int_0^\infty \int_0^\infty \!
  d\sigma_1 d\sigma_2 D_{\mu\nu} (z_{12}) ]^{-1}}{[1 + (ig)^2  v_1^\mu v_1^\nu
  \int \int\!
  d\sigma_1 d\sigma_2 D_{\mu\nu} (z_{1})]^{-1} [1 + (ig)^2  v_2^\mu v_2^\nu
  \int \int \!
  d\sigma_1 d\sigma_2 D_{\mu\nu} (z_{2}) ]^{-1}} - 1
   \ ,
\label{eq:lo_angle}
\end{eqnarray}
where $D_{\mu\nu} (z)$ is the free gauge-field propagator and
$z_{12}^\rho = (v_{1}^\rho \sigma_1 \pm v_{2}^\rho \sigma_2 \pm b^\rho)$,
with $b_\rho$ being purely transverse, i.e.,
$ b_\rho = (0,0, \mathbf b_\perp)$, so that
$z_{1,2}^\rho = (v_{1,2}^\rho \sigma_1 \pm v_{1,2}^\rho \sigma_2)$.
Employing a covariant gauge, it is convenient to express the gluon
propagator in the following general form
(tacitly performing a dimensional regularization of the UV divergences in
terms of $\omega = 4 - 2 \epsilon$ and by using the auxiliary
infrared (IR)
cutoff $\lambda^2$)
\begin{equation}
  D_{\mu\nu} (z)
=
    g_{\mu\nu} \partial^2 D_1 (z^2)
  - \partial_\mu \partial_\nu D_2 (z^2) \ ,
\label{prop_derivative}
\end{equation}
where
\begin{equation}
  \partial^2
=
  2 \omega \partial + 4 z^2 \partial^2 , \
  \partial_\mu \partial_\nu
=
  2 g_{\mu\nu} \partial
  + 4 z_\mu z_\nu \partial^2 , \ \partial \equiv \partial_{z^2} \ .
\label{eq:partial}
\end{equation}
The above representation allows in Eq.\ (\ref{eq:lo_angle}) the separate
calculation of the integrals over $\sigma_1$ and $\sigma_2$
by carrying out a Laplace transformation of the functions $D_{1,2}$ and
their derivatives $D_{1,2}^{(k)}$:
\begin{eqnarray}
  D_{1,2}^{(k)} (z^2; \epsilon, \mu^2, \lambda^2)
=
  (-1)^k \int_0^\infty\! d\alpha \alpha^k
  {\rm e}^{- \alpha z^2} \
  \tilde D_{1,2} (\alpha; \epsilon, \mu^2, \lambda^2) \ .
\label{eq:prop_laplace}
\end{eqnarray}
Hence, the angular dependence stems solely from the integral
\begin{equation}
  \int_0^\infty \int_0^\infty \!
  d\sigma_1 d\sigma_2
  {\rm e}^{- \alpha z_\mu z^\mu}
  =
  \int_0^\infty \int_0^\infty \!
  d\sigma_1 d\sigma_2
  {\rm e}^{- \alpha (v_{1} \sigma_1 - v_{2} \sigma_2 - b)^2} \ .
\label{eq:basic_int_1}
\end{equation}
In general, the scalar products $(v_{1,2}\cdot b)$ are not vanishing,
so that the result of the integration in
Eq.\ (\ref{eq:basic_int_1}) is complicated.
However, for our purposes it suffices to evaluate this
expression using the following approximations:
First, we employ the Feynman gauge and consider the
gauge field as being Abelian, given that at the order $g^2$ of the
expansion the nonlinear terms in the gluon field do not contribute.
This gives
\begin{equation}
  D_{\mu\nu} (z)
=
  -i g_{\mu\nu}
  \int\! \frac{d^4 k}{(2\pi)^4}
  \frac{{\rm e}^{- ik \cdot z}}{k^2 - \lambda^2 +i0}
\label{eq:D-mu-nu}
\end{equation}
with the result that in Eq.\ (\ref{prop_derivative}) $D_2 = 0$.
Next, we restrict our attention to the case of collinear
kinematics, i.e., when the transverse distance $\mathbf b_\perp$
equals zero, appealing to the fact that worldlines are collimated
within a very narrow conus with $\Delta\eta$ values in the range
$2.0<|\Delta\eta|<4.8$.
In this case, the angular dependence stems only from
Eq.\ (\ref{eq:basic_int_1}), recalling that
$v_{1,2}^2 = 1$, $\mathbf b_\perp = 0$ and defining
\begin{eqnarray}
A (Y_1, Y_2, \phi_1, \phi_2) = (v_1\cdot v_2) =
\frac{1}{m^2} \left(
m_T^2 \ \cosh \Delta Y - p_\perp^2 \ \cos \Delta \phi \right) \ ,
\label{eq:A}
\end{eqnarray}
where $\Delta \phi = \phi_1 - \phi_2$, and $\Delta Y = Y_1 - Y_2$.
\begin{figure}[t]
\centering
\centerline{\includegraphics[scale=0.45,angle=0]{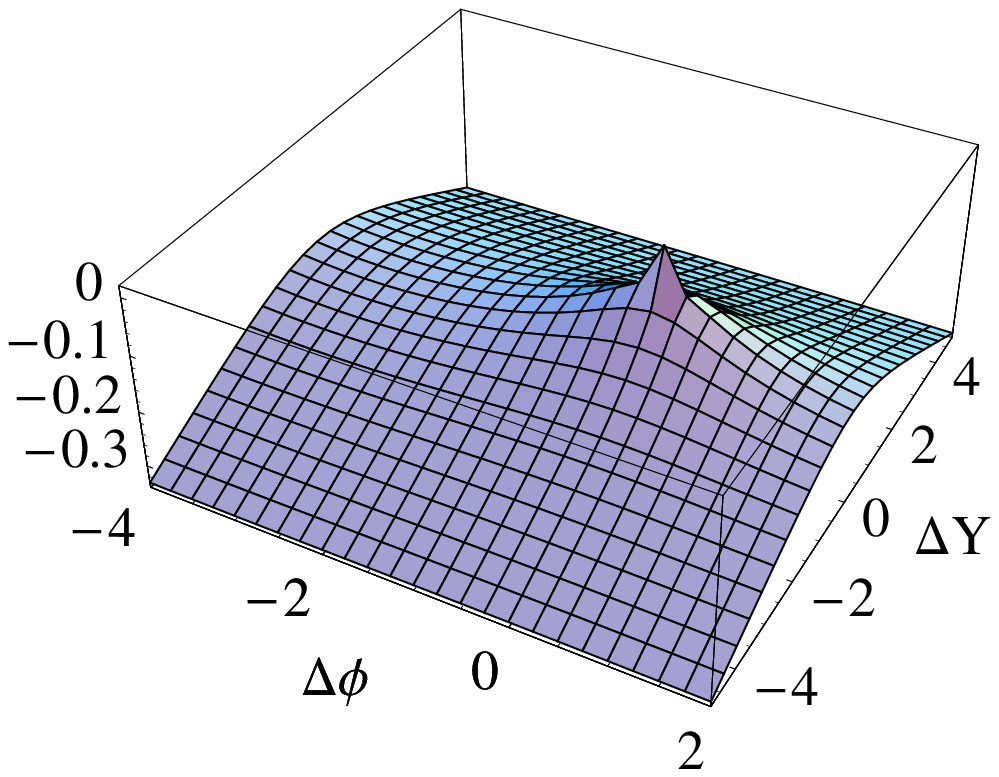}~~~~~~~~
~~~~~~~~~~~~\includegraphics[scale=0.45,angle=0]{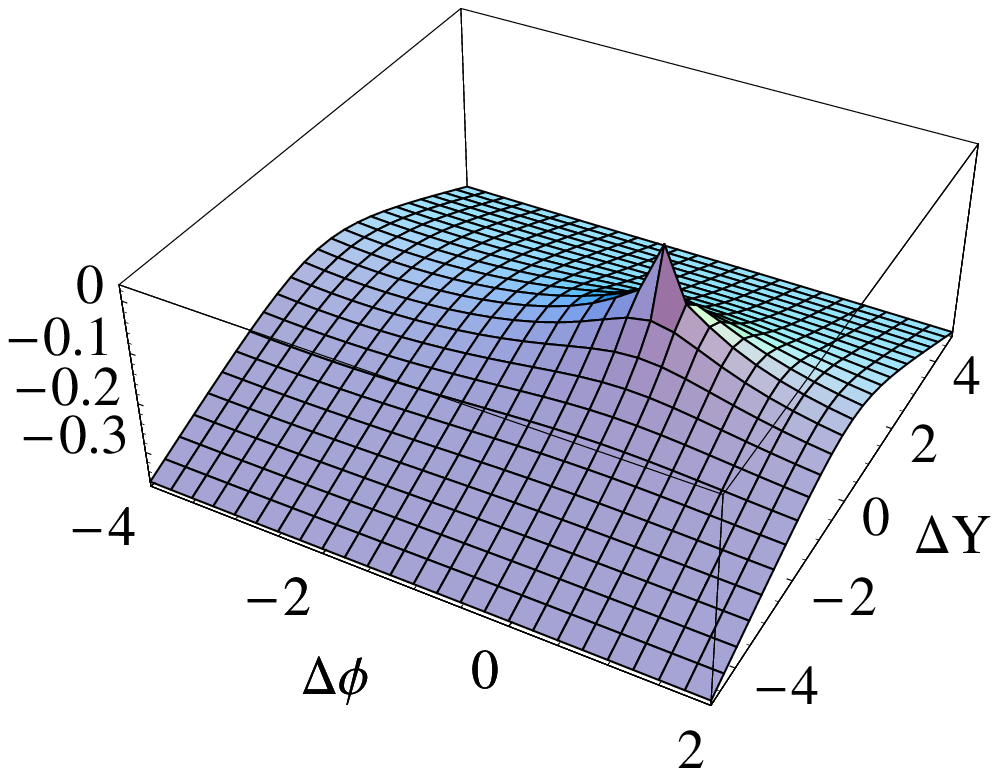}~~~~~~~~
~~~~~~~~~~~~\includegraphics[scale=0.45,angle=0]{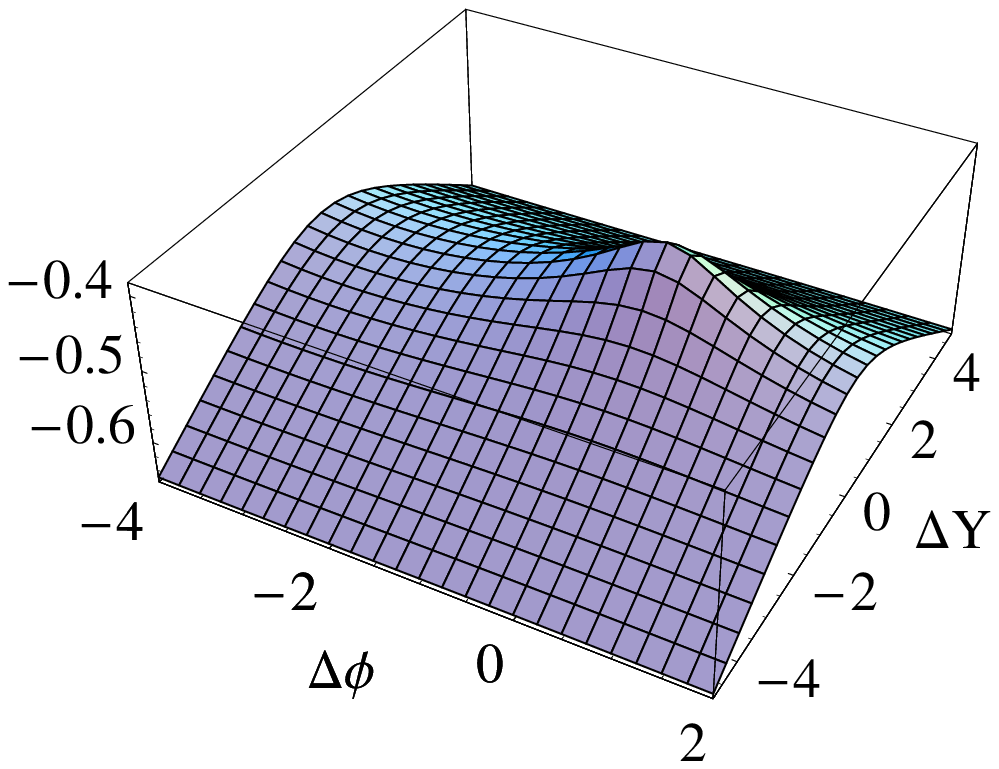}~~~~~~~~
}
\caption{2-D representation of the leading-order angular part
$K^{(1)}(\Delta Y, \Delta\phi)$ of the total two-particle
correlation function---given by Eq. (\ref{eq:corr_soft})---in
terms of interacting Wilson (world-) lines vs. $\Delta Y$ and
$\Delta\phi$ for (from left to right)
$[m_T = 3.5, p_\perp = 3.4], \ [m_T = 5.0, p_\perp = 4.9]$, and
$[m_T = 150.0, p_\perp = 140.0]$ in units of MeV.
The strong coupling constant and the UV- and IR-regulators are
fixed to give $\alpha_s \ln \mu^2/\lambda^2 = 0.1$.\\
\label{fig:angular}}
\end{figure}

It is remarkable that $A$ (and, as a consequence, the whole
function $K^{(1)}$) depends only on the {\it differences} of
the rapidities and the azimuthal angles of the
produced particles, so that we do not have to average over
their sums.
Hence, we have
$$
A (Y_1, Y_2, \phi_1, \phi_2) =
A (\Delta Y, \Delta \phi) \ ,
$$
which represents a major advantage of parametrization
(\ref{eq:parameter_rapidity}).
Then, we get
\begin{eqnarray}
  \int_0^\infty \int_0^\infty \! d\sigma_1 d\sigma_2
  {\rm e}^{- \alpha (\sigma_1^2 +\sigma_2^2 \pm 2 A \sigma_1 \sigma_2)}
=
  \frac{\pm \pi \mp 2
  \arctan\sqrt{\frac{A^2}{1- A^2}}}{4\alpha \sqrt{1-A^2}} \ ,
\label{eq:basic_int_2}
\end{eqnarray}
while the UV and the IR singularities originate from the
$\alpha$-integration in Eq.\ (\ref{eq:prop_laplace}) and have been
separated out from the angular part.

Therefore, performing the integrations in (\ref{eq:lo_angle}),
one finds after renormalization the following expression
for the angular-dependent part
\begin{eqnarray}
  K^{(1)}(\Delta Y, \Delta \phi; p_\perp, m_T)
=
  \frac{\alpha_s \ C_{\rm F}}{2\pi} \ln \frac{\mu^2}{\lambda^2} \
  \left[ {A} \cdot
  \frac{\pm \pi \mp 2
  \arctan\sqrt{\frac{A^2}{1- A^2}}}{\sqrt{1-A^2}} - 1\right]
\label{eq:corr_angle}
\end{eqnarray}
in terms of the differences of the azimuthal angles and
pseudorapidities, where $A$ is defined in Eq. (\ref{eq:A})
and $\mu$ and $\lambda$ are the UV- and IR-cutoff scales,
respectively.
The result for the angular part as a function of the
absolute value $|\Delta Y|$ is shown graphically in Fig.\ \ref{fig:angular}
for different values of the parameters $m_T$ and
$p_\perp$.

\begin{figure}[h]
\centering
\centerline{\includegraphics[scale=0.6,angle=0]{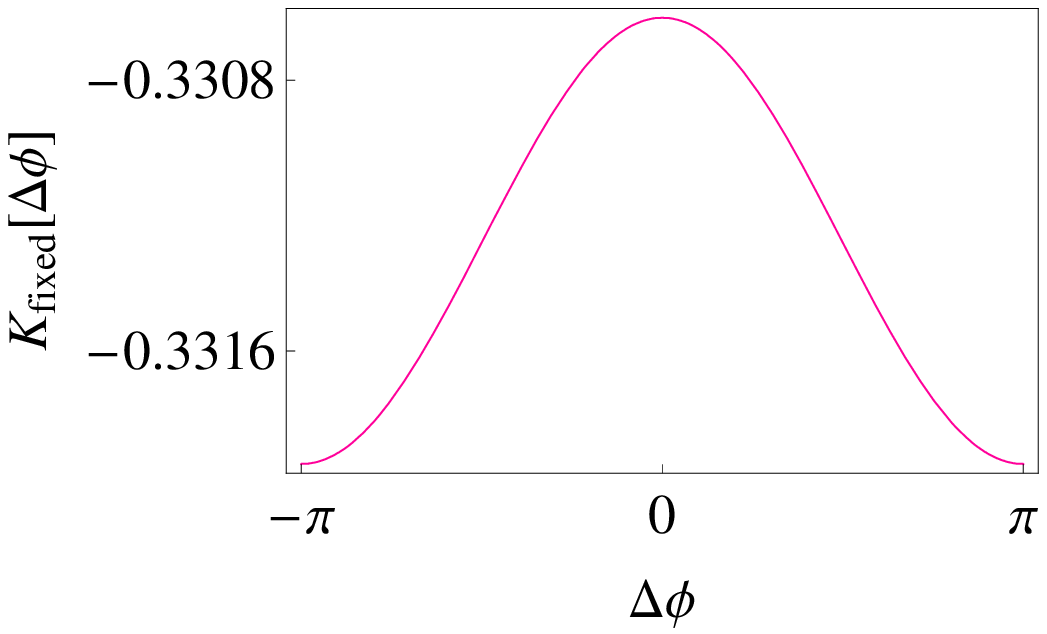}~~~~~~
~~~~~~~~~~~~\includegraphics[scale=0.6,angle=0]{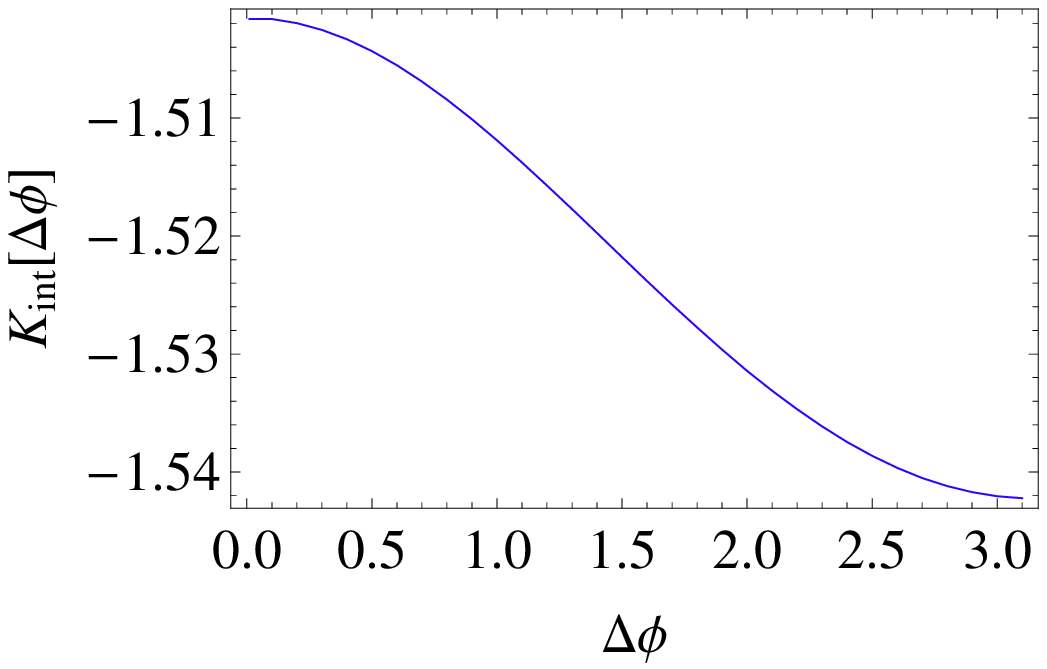}}
\caption{1-D soft correlation functions.
Left panel: $K (\Delta Y, \Delta \phi)$ for fixed rapidity at the
endpoint of the acceptance region $\Delta Y = 4.8 $ with a
flat maximum at near side.
Right panel: Integrated 2-D soft correlation function
$K (\Delta \phi) = \int_2^{4.8} K (\Delta Y, \Delta \phi) \ d \Delta Y$
for $m_T = 3.5, \ p_\perp = 3.4$.
The signature of the ridge is the smooth maximum at
$\Delta \phi \approx 0$.
\label{fig:rapidity}}
\end{figure}

In order to make the ridge-like structure more visible, we
first fix the value of $\Delta Y$ and display the 1-D--``slice''
of the 2-D function $K(\Delta Y = 4.8, \Delta \phi)$
in the left panel of Fig. \ref{fig:rapidity}.
Following a similar procedure to that in Ref.\ \cite{CMS_2010_ridge},
we average the two-particle azimuthal correlation function over
$|\Delta Y|$ from $2.0$ to $4.8$ and plot the 1-D correlation function
vs. $\Delta\phi$ in the right panel of Fig. \ref{fig:rapidity}.
It can be compared with Fig. 8 (last 2 columns) given by the CMS
Collaboration in Ref.\ \cite{CMS_2010_ridge}.

The main observation from Fig. \ref{fig:angular} is that the
soft correlation function given by Eq.\ \ref{eq:corr_soft} reproduces the
gross features of the near-side ridge structure found by the CMS
Collaboration \cite{CMS_2010_ridge}.
In particular, it depends explicitly on the differences in rapidity and
azimuthal angle between pairs of produced particles, as well as on the
transverse momentum, and exhibits a localized maximum at
$\Delta\phi\approx 0$ that extends to larger values of $\Delta Y$.
However, it is much less pronounced than the experimental finding,
as it also becomes clear from inspection of the 1-D correlation
$K[\Delta\phi]$ in Fig.\ \ref{fig:rapidity} which peaks at
$\Delta \phi \approx 0$, decreasing significantly at
$\Delta \phi \approx \pi$.
The collimation in the near-side region is due to the
$\cos \Delta \phi$ term and originates from the mutual interactions
of the Wilson exponentials evaluated along the worldlines of the
produced particles.

\section{Summary of results and discussion}
\label{sec:summary}

Let us summarize our results and discuss them further.
\begin{itemize}
\item
In this work, we discussed the possibility of quantum entanglement
for pairs of charged particles produced in high-energy
$pp$-collisions at the LHC.
We argued that the long-range correlations observed in such collisions
in the CMS experiment can be assumed to arise from quantum entangled
states of Wilson lines that inevitably enter the description of
fragmentation functions.

\item
Using methods and ingredients of gauge field theory, as it applies
to unintegrated parton distribution or fragmentation functions within
transverse-momentum dependent factorization schemes, we calculated this
correlation in terms of interacting Wilson lines in leading order of
the coupling constant.
Note, however, that the structure of the non-Abelian gauge links in the
operator definitions of the TMD PDF/PFFs is quite complicated
(see, e.g., \cite{BM07}) and still a matter of dispute
\cite{CKS_2010,Col11}.
In the presented investigation, we have simplified significantly our
task by taking into account only the one-gluon interactions of two
semi-infinite Wilson lines.
Moreover, we did not take into account the gluon TMDs and ignored
evolution effects both in rapidity and the momentum.

\item
The principal result of our calculation is that the long-range
near-side angular correlation has a pattern in the
$\Delta Y-\Delta\phi$ space that bears the main characteristics
of the near-side ridge observed by the CMS Collaboration.
To be specific, the angular part of the correlation function
(see Fig.\ \ref{fig:angular}) shows two distinct regions: a near-side
localized ridge around $(\Delta Y, \Delta\phi)\approx 0$ and an
away-side elongated ridge extending in $\Delta Y$ up to five units.
The near-side ridge is much less prominent than the observed sharp
peak in the CMS data.
This may indicate that parton interactions with the medium and
other collective effects may play an important role in further
enhancing the particle correlations to the observed size of the
ridge.
However, overall, the angular structure of the computed correlation
emulates the main characteristics of the near-side ridge
qualitatively.
The ridge structure is also observed in the 1-D azimuthal
angular correlation averaged over $\Delta Y$ in the range
$[2 \div 4.8]$ with a shallow peak around $\Delta\phi \approx 0$.

\item
Given the high multiplicity of $N\approx 110$ in the CMS data, one may
ask about the role played by multiparton correlations.
This issue was addressed in \cite{CMS_2010_ridge} via the
zero-yield-at-minimum (ZYAM) method by calculating the associated
yield, i.e., the number of other particles correlated with a specific
particle.
Such a detailed analysis is outside the scope of the present
investigation.
However, as far as multi-particle entangled states are concerned,
it was shown quite recently \cite{Qbit_2011} that the decay
rate of the so-called Greenberger-Horne-Zeilinger quantum states (which
would resemble multi-particle entangled states in our worldline
approach) increases quadratically with the number of particles $N$,
thus entailing the contribution of correlations due to multiple
sources to be suppressed in the relatively small rapidity range
probed in the CMS experiment.

\item
In this work, we left the ground state in the considered correlators
[cf.\ Eqs.\ (\ref{eq:product_1}) and (\ref{eq:product_2})] unspecified.
In a more realistic treatment, the ground state
$\left. |{\rm GS}\right \rangle$,
which enters the definitions of the soft factors, can be conceived of
as being either the nonperturbative QCD vacuum, or, say, the color-glass
condensate, etc.
Therefore, our approach seems to be more akin to the Wilson-line
representation of high-energy collisions
(see e.g., \cite{Wilson_HE} and Refs.\ cited therein) and less similar
to the gluon-saturation picture.
\end{itemize}

\section{Conclusions}
\label{sec:concl}

In conclusion, using a theoretical framework which emulates particle
tracks produced in high-energy $pp$-collisions at the LHC by
interacting Wilson lines, we discussed the possibility of quantum
entanglement for pairs of charged particles.
This soft correlation is universal and leads to 2-D and 1-D
two-particle angular correlation functions which reproduce the
characteristic ridge structure observed in the CMS experiment---though
in less pronounced form.
Hadronic properties may be taken into account in a refined analysis
to improve the agreement with the data more quantitatively.

\acknowledgments
I.O.C. thanks the Organizers of the Conference ISMD 2010 (Antwerp)
for the invitation and support, where he first learned about the CMS
result.
He also thanks X. Janssen and the participants of his seminars in
Antwerp and Li\`ege for useful discussions.

\end{document}